\documentstyle[twoside,fleqn,espcrc2]{article}

\newcommand{\AmS}{{\protect\the\textfont2
  A\kern-.1667em\lower.5ex\hbox{M}\kern-.125emS}}

\newcommand{\be}{\begin{equation}}
\newcommand{\ee}{\end{equation}}
\newcommand{\bea}{\begin{eqnarray}}
\newcommand{\eea}{\end{eqnarray}}
\newcommand{\dr} {$ \Delta\rho $}
\newcommand{\nn} {\nonumber}
\newcommand{\half}{\frac12}

\title{Monte-Carlo study of the $\Delta\rho$ parameter\thanks{Presented
by Ph.\ Boucaud ({\tt phi@qcd.th.u-psud.fr}).
Work supported by EC contract ERBCHBICT941067,
by DGICYT project AEN 94-218, by Acci\'on Integrada 
Hispano-Francesa HF94-150B and HF95-296B and by Caja de Ahorros de la
Inmaculada.}}

\author{J.L. Alonso\address{Departamento de F\'{\i}sica Te\'orica, 
        Universidad de Zaragoza, 50009 Zaragoza, Spain}, 
        Ph.\ Boucaud\address{LPTHE, Universit\'e Paris XI,
         91405 Orsay Cedex, France}
        and A.J. van der Sijs$^{\rm a}$}

\begin{document}

\begin{abstract}
We present  results concerning a lattice study of the electroweak 
$\rho$-parameter. We have used an SU(2)$\,\times\,$U(1) symmetric chiral
Yukawa model built with Zaragoza 
fermions. The decoupling of the species doublers in this model is verified
numerically.
We find that the numerical data for \dr\ are well described by one-loop 
perturbation theory 
in the same finite volume and with the same finite cut-off.
However, a finite cut-off can cause substantial deviations of \dr\ from
the standard value, even in infinite volume.

\end{abstract}

\maketitle

\section{INTRODUCTION}
An important quantity in the Standard Model of electroweak interactions
is the  $\rho$-parameter, defined as the relative strength 
between the neutral and charged current interactions at zero momentum transfer.
At tree level at low momentum, the electroweak interaction is described by the 
standard  current-current interaction and  it is well known that
$\rho_{\rm tree} = M_W^2 / M_Z^2 \cos^2\theta_W = 1.$
When quantum effects are taken into account, 
$\rho = ( 1 - \Delta\rho)^{-1}$ deviates from 1. 
We will be concerned here only with the dominant ``universal" 
(i.e.\ process-independent) part of \dr\ coming from the 
difference in
vacuum polarization of the $W$ and $Z$ propagators.

In the Standard Model, the dominant contribution comes from 
the $t$-$b$ mass splitting.
In the limit $m_b^2\ll m_t^2$ it is given by
\be 
\Delta\rho \ =\ N_c {\sqrt{2} \, G_F \over 16 \pi^2} m_t^2
\ =\ N_c {y^2 \over 16 \pi^2} \, .
\label{deltarhostandard}
\ee

 From a phenomenological point of view, this quantity is an interesting
tool to  investigate the existence of unknown physics through loop effects.
This is exemplified by the bounds put on the top quark mass before its discovery.
As it appears in eq.~(\ref{deltarhostandard}), there is a strong dependence on the 
mass of a heavy fermion running in the loop. In particular such a fermion   
does not decouple from the low-energy physics. 
It would be desirable to control the persistence or not of this non-decoupling 
phenomenon non-perturbatively, for large values of the mass, and to study 
its relation with the issue of triviality. 
This problem has been
addressed in a large-$N_F$ approximation before \cite{santi}.

 From the lattice point of view, these characteristics have some relevance
to the doubling problem. 
We know for example that heavy doublers treated \`a la Wilson-Yukawa
do not decouple in the continuum limit. 
Through the relation (\ref{deltarhostandard}), 
\dr\ can be used as a counter of fermions. 
It thus provides us with a tool to test 
models of chiral lattice fermions.

\section{MEASURING \dr\ ON THE LATTICE}
A numerical  prediction for \dr\  from a direct study 
of the gauge boson propagators would require 
a tractable lattice formulation of a four-dimensional chiral gauge theory. 
This task is beyond our  present capabilities. 
However, the gauge interactions are weak compared  to the Yukawa and scalar 
ones and it is a reasonable approximation to neglect all the gauge couplings. 
There are Ward identities relating the vacuum 
polarization tensors of the gauge bosons to 
those of the charged and neutral goldstone particles in the fermion-scalar 
sector, and $\rho$ can be expressed as the ratio of their wave function
renormalization constants:
$ \rho \ =\ Z_0 / Z_+$
\cite{Lytel,Barbieri}.

In the gaugeless approximation, therefore,
only the scalar-fermion sector of the Standard Model needs to be latticized
to obtain $\rho$.
For this purpose, we have used an SU(2)$\,\times\,$U(1) symmetric chiral Yukawa 
fermion model built with Zaragoza fermions. The lattice action is:
\begin{eqnarray}
\textstyle
{\cal S} = 
 -\kappa \sum_{x,\mu} & \!\!\!\!\!\!
      {\rm Tr} \left[ \Phi^+(x) 
\left( {\Phi(x+\hat\mu) + \Phi(x-\hat\mu) \over 2}
       \right)\right] \hspace {.5 cm} & \!  \nn  \\ 
\textstyle
  + \sum_{x,\mu}  & \!\!\!\!\!\!
          \bar\psi (x)\gamma_\mu 
        {\psi (x+\hat\mu)- \psi (x-\hat\mu) \over 2} 
\hspace {1.6 cm} &  \!  \label{action}   \\ 
\textstyle
  + \sum_{x} \hspace {.29 cm} & \!\!\!\!\!\!
          \left(\bar\psi^{(1)}(x) \Phi(x) Y P_R   
\psi^{(1)}(x) + h.c. \right)  \!   \nn
           \, . &
\end{eqnarray}
$\Phi$ is a fixed modulus Higgs field ($\Phi \in$ SU(2)),
$\psi$ is a  fermionic field for two fermion doublets, $P_R$ is the
right projector $\half(1 + \gamma_5)$ and
$Y = \left(\matrix { y \; 0 \atop 0 \; 0}\right)$  
is the matrix of Yukawa couplings,
where $y=y_{\rm top}$ and $y_{\rm bottom}$ is set to zero.
The fermion field $\psi^{(1)}(x)$ is the average of the $\psi$ fields over the vertices
of the hypercube at $x$ \cite{Zaragoza}; in momentum space it can be written
as $\psi^{(1)}(p) = F(p) \psi(p)$ with ``form factor''
$F(p) = \prod_\lambda \cos(a p_\lambda / 2)$.  
This model is expected to describe two coupled fermion doublets and thirty  
massless and uncoupled doubler doublets \cite{Zaragoza}.
We have simulated the action (\ref{action}) with  
the Hybrid Monte-Carlo algorithm  
(the reason why we have doubled the fermion content) 
and present results for   
$6^3\times 12$ and $8^3\times 16$ lattices. 
Details will be given in \cite{inprepar}.

\section{DECOUPLING OF THE DOUBLERS}
With Zarogoza fermions, the decoupling of the doublers is known 
to hold in perturbation theory \cite{Zaragoza}.
Numerically, the structure of the phase diagram 
was previously found in good agreement with a mean-field prediction, 
giving some indication that the decoupling works \cite{phasediag}. 
We have done  some  further tests reported here:

\noindent$\bullet$ 
 From the three point function
$G_3(k,p,q) =  \sum_{x,y,z}
\langle \Pi_0(x) \psi(y) \bar\psi(z) \rangle \exp[-i (k.x + p.y - q.z)] $,
where $\Pi_0(x) = $ Tr $\left[\Phi(x) \sigma_3 \right]$ 
is an interpolating field for the neutral goldstone, we have extracted
the renormalized three-point couplings $y_{\rm 3pt}$ for the usual fermion
(i.e. with momentum near $p=0$), and $y_{\rm db}$ for a doubler
(i.e. with momentum near $p_\mu=q_\mu=\pi a^{-1}$) 
respectively. We find as expected that $y_{\rm db} 
\approx y_{\rm 3pt} * F^2$, 
i.e.\ the coupling of the doublers is suppressed  by the form factor $F^2$.
This suppression factor is
around $10^{-2}$ for the minimum momentum allowed for a fermion on an
$8^3\times 16$ lattice and will decrease if the volume is increased.

\begin{figure}[htb]
\vspace {4.0cm}
\includegraphics{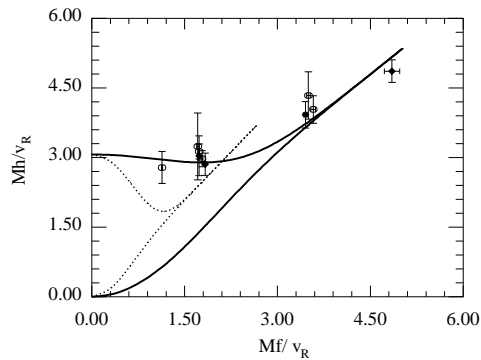}
\caption{ 
Higgs mass as a function of the fermion mass in units of 
$v_r$.
MC data for $6^3\times 12$ and $8^3\times 16$ lattices are shown,
and the solid line is the perturbative result with 2 doublets,
for a scale ratio $\log (v_r / \Lambda ) = - 4/3$.
The analogous result for 32 doublets is given by the dotted curves.
 }
\label{fig:higgsmasse}
\end{figure}

\noindent$\bullet$ 
We have studied  the one-loop renormalization group equations for the evolution
of the scalar and Yukawa coupling constants for the continuum model with 2
and 32 fermions doublets and compare the solutions with our Monte-Carlo data 
in fig.~\ref{fig:higgsmasse}. The data seem to follow the one-loop
prediction and support the expectation that we have only two dynamical doublets,
without coupled doublers.
 
\noindent$\bullet$
We have compared  results obtained for \dr\ 
with Zaragoza  and  naive fermions for two points in the phase diagrams
with comparable fermion mass in lattice units.
We defined  the renormalized coupling constant, $y_r$, as the ratio of the renormalized
fermion mass divided by the renormalized vacuum expectation value of the 
scalar field, $v_r$.
On a $6^3\times 12$ lattice, the comparison between Monte-Carlo data and perturbation
theory  gives
$\left[\Delta\rho_{\rm MC}/y_r^2\right] /\left[\Delta\rho_{\rm pert}/y^2\right]$ = 0.98(13)
for $\kappa$ = 0.24, $y$=0.6, $a m_f$=0.311 in the naive case
and 
$\left[\Delta\rho_{\rm MC}/y_r^2\right] /\left[\Delta\rho_{\rm pert}/y^2\right]$ = 0.92(18) 
for $\kappa$ = 0.31, $y$=1.0, $a m_f$=0.345 in the Zaragoza case. 
This means that for both fermion models, the numerical results are well 
compatible with their respective perturbative predictions. On the other hand, 
perturbation theory predicts  that
$\Delta\rho_{\rm naive} / \Delta\rho_{\rm Zaragoza}$ equals  17.13 and 17.26 
for $a m_f$ equal to 0.311 and 0.345 respectively (this ratio becoming 16 
in the infinite volume limit when $a m_f \ll 1$). 
This is a clear indication that the Zaragoza doublers 
are decoupled and do not contribute to \dr.

\section{RESULTS FOR \dr}
\begin{figure}[htb]
\vspace{9pt}\vspace {4.6cm}
\includegraphics{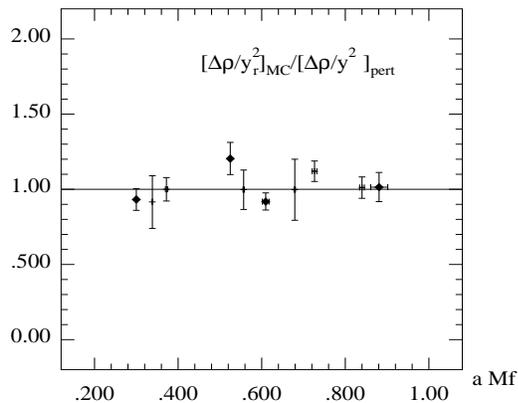}
\caption{ 
Ratio of  the Monte-Carlo $\Delta\rho / y_r^2$ 
divided by the finite-volume, finite-cut-off perturbative value.
Data from $6^3\times 12$ and $8^3\times 16$ lattices are shown.
}
\label{fig:deltarho}
\end{figure}

To detect possible non-perturbative effects, we have 
compared
the results obtained from the simulations with 
the one-loop perturbative predictions  {\bf at finite volume \boldmath {$V$}
and finite cut-off \boldmath{$a^{-1}$}}. 
Namely, for each lattice ($\kappa,\,y,\,V$) we have extracted  
from the numerical data the value of the fermion mass in lattice units, 
$a m_f$,  and used it  in a finite-volume fermionic one-loop  perturbative 
computation of \dr.
Fig.~\ref{fig:deltarho} shows the ratio of the Monte-Carlo value
$\Delta\rho_{\rm MC}/y_r^2$ divided by the corresponding perturbative 
prediction.
The numerical  results seem to be  well represented  
by the finite-volume, finite-cut-off one-loop perturbative calculation
for both volumes and all the values of $y$ so far explored.

\section{CONCLUSIONS}
\begin{figure}[htb]
\vspace{9pt}\vspace {3.8 cm}
\includegraphics{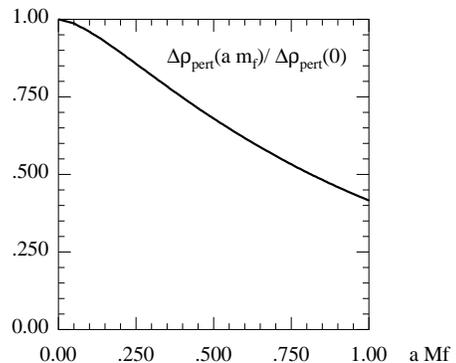}
\caption{ 
Finite-cut-off effects in $\Delta\rho$. We plot the 
perturbative one-loop estimate for the ratio
$\Delta\rho (a m_f) / \Delta\rho (0)$ 
obtained with  the action (2) at infinite volume. 
\hfill\break\hfill\break\hfill\break
}
\label{fig:nombredefermion}
\end{figure}

The dominant part of \dr\ can be computed within the scalar-fermion
sector of the Standard Model. 
This can be done on the lattice with present techniques.
We have found different kinds of numerical evidence that the 
decoupling mechanism
works as expected with the Zaragoza proposal for lattice fermions.
\dr\ is well described by the perturbative prediction. 
Nevertheless, even if \dr\ is always given by perturbation theory, 
cut-off effects can  be important \cite{santi}. 
When the mass of the fermion approaches the cut-off, \dr\ deviates from the
standard result  given in  eq.~(\ref{deltarhostandard}), 
valid when $m_f /\Lambda \ll 1$. 
In fig.~\ref{fig:nombredefermion} we give the ratio 
$\Delta\rho_{\rm pert}(a m_f) /\Delta\rho_{\rm pert}(0)$ 
for the lattice model we have used here.
It shows that cut-off effects in   \dr\ can be quite substantial.
\vskip -1.62 cm
A more detailed account of this work, including a more complete list 
of references and acknowledgements, will appear soon \cite{inprepar}.

\end{document}